\documentclass[a4paper,11pt]{article}

\usepackage{balance}
\usepackage[font={small}, labelsep={period},labelfont={bf}]{caption}

\usepackage[utf8]{inputenc}
\usepackage[T1]{fontenc}
\usepackage{graphicx}
\usepackage{epsfig}
\usepackage[english]{babel}
\usepackage{listings}
\usepackage{hyperref}
\usepackage{authblk}
\usepackage{fancyhdr}
\usepackage{floatflt}
\usepackage{float}
\usepackage{wrapfig}
\usepackage{rotating}
\usepackage{longtable}
\usepackage{multirow}
\usepackage{adjustbox}
\usepackage{makecell}
\usepackage{chngcntr}
\usepackage{relsize}
\usepackage{amsmath}
\usepackage{amsfonts}
\usepackage{amssymb}
\usepackage[amssymb, cdot, mediumqspace, thickqspace]{SIunits}
\usepackage[a4paper,top=3cm,bottom=3cm,left=2.5cm,right=2.5cm]{geometry}
\usepackage{mathrsfs}
\usepackage{booktabs}
\usepackage{braket}
\usepackage{subfigure}
\usepackage[titletoc, title]{appendix}
\usepackage{pdfpages}

\usepackage[style=phys, biblabel=brackets, citestyle=numeric-comp, backend=biber, sorting=none, doi=true, url=true, eprint=true, maxbibnames=2]{biblatex}

\usepackage[style=phys, biblabel=brackets, citestyle=numeric-comp, backend=biber, sorting=none, doi=true, url=true, eprint=true, maxbibnames=2]{biblatex}

\DeclareFieldFormat{titlecase}{#1} 
\DeclareFieldFormat{doi/url-link}{#1} 
\DeclareSourcemap{ 
  \maps[datatype=bibtex]{
    \map[overwrite]{
      \step[fieldsource=doi, final]
      \step[fieldset=eprint, null]
      \step[fieldset=url, null]
    }  
  }
}

\AtEveryBibitem{
    \clearfield{month}
}

\DeclareSourcemap{
 \maps[datatype=bibtex,overwrite=true]{
  \map{
   \pernottype{inproceeding}
    \step[fieldsource=Collaboration, final=true]
    \step[fieldset=usera, origfieldval, final=true]
  }
 }
}

\renewbibmacro*{note+pages}{
  \printfield{note}%
  \setunit{\bibpagespunct}%
  \printfield{pages}%
  \iffieldundef{pages}
    {%
      \clearfield{doi}%
    }%
    {%
    }%
}

\usepackage{placeins}
\usepackage{titlesec}
\usepackage{mathtools}
\usepackage{enumitem}
\usepackage{color}
\usepackage{mhchem}
\usepackage{slashed}
\usepackage{feynmp}
\usepackage{feynmp-auto}
\usepackage{lineno}
\usepackage[bottom]{footmisc}
\usepackage{tabularx}
\usepackage{lineno}
\usepackage{multirow}
\usepackage{tikz} 
\usetikzlibrary{shapes,arrows}
\usepackage{fancyvrb}
\usepackage{lscape}
\usepackage{comment}
\usepackage{cleveref}
\usepackage{orcidlink}
\usepackage{authblk}

\definecolor{denim}{rgb}{0.08, 0.38, 0.74}
\definecolor{darkblue}{rgb}{0.00, 0.00, 0.5}
\definecolor{internationalkleinblue}{rgb}{0.0, 0.18, 0.65}
\definecolor{bondiblue}{rgb}{0.0, 0.58, 0.71}
\definecolor{blue_ryb}{rgb}{0.01, 0.28, 1.0}
\definecolor{cobalt}{rgb}{0.0, 0.28, 0.67}
\definecolor{lightblue}{rgb}{0.145,0.6666,1}

\newcommand{\lphi}{$\tau^- \rightarrow \ell^-\phi$}
\newcommand{\ephi}{$\tau^- \rightarrow e^-\phi$}
\newcommand{\mphi}{$\tau^- \rightarrow \mu^-\phi$}
\newcommand{\phipi}{$D^{+}_\mathrm{s} \rightarrow \phi\pi^{+}$}
\newcommand{\threepi}{$\tau^- \to \pi^-\pi^+\pi^-\nu_\tau$}

\newcommand {\emep} {$e^{+}e^{-}$}
\newcommand {\deltae} {$\Delta E_{\tau}$}
\newcommand {\invmass} {$M_{\tau} $}

\newcommand {\invfem}{~fb$^{-1}$}
\newcommand {\invapto}{~ab$^{-1}$}

\newcommand{\gevc}{GeV/$c$}

\titleformat{\paragraph}
{\normalfont\normalsize\bfseries}{\theparagraph}{1em}{}
\titlespacing*{\paragraph}
{0pt}{3.25ex plus 1ex minus .2ex}{1.5ex plus .2ex}

\fancypagestyle{plain}{%
  \fancyhf{}\fancyfoot[C]{ \thepage}%
}

\hypersetup{
	colorlinks = true,
	allcolors = internationalkleinblue,
	linktocpage
}

\graphicspath{ {./figures/} }

\bibliography{bibliografia}

\begin{document}

\begin{figure}
\includegraphics[width=0.25\textwidth]{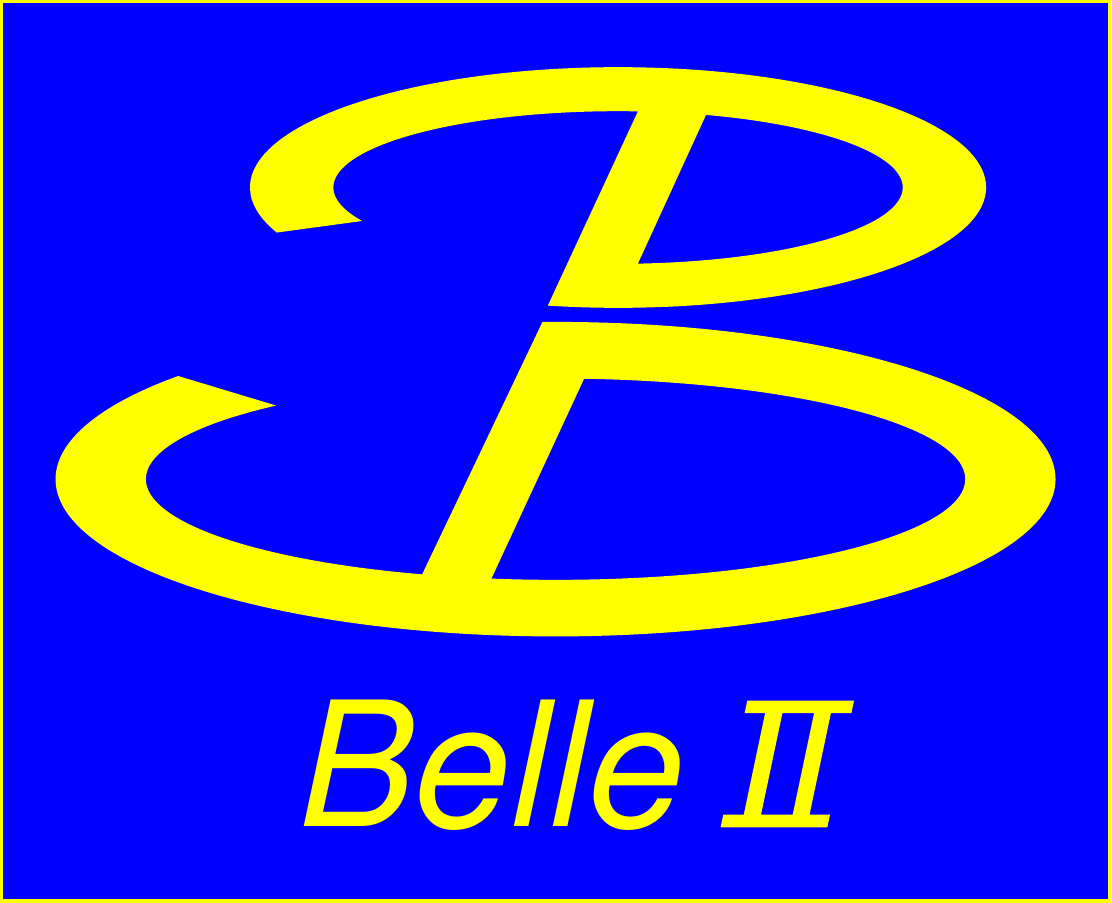}
\vspace{-3 cm}
\end{figure}

\begin{flushright}BELLE2-CONF-2023-004\\
\vspace{2 cm}
\end{flushright}

  \author{F.~Abudin{\'e}n\,\orcidlink{0000-0002-6737-3528}} 
  \author{I.~Adachi\,\orcidlink{0000-0003-2287-0173}} 
  \author{K.~Adamczyk\,\orcidlink{0000-0001-6208-0876}} 
  \author{L.~Aggarwal\,\orcidlink{0000-0002-0909-7537}} 
  \author{P.~Ahlburg\,\orcidlink{0000-0002-9832-7604}} 
  \author{H.~Ahmed\,\orcidlink{0000-0003-3976-7498}} 
  \author{J.~K.~Ahn\,\orcidlink{0000-0002-5795-2243}} 
  \author{H.~Aihara\,\orcidlink{0000-0002-1907-5964}} 
  \author{N.~Akopov\,\orcidlink{0000-0002-4425-2096}} 
  \author{A.~Aloisio\,\orcidlink{0000-0002-3883-6693}} 
% ...
\affil{(The Belle~II Collaboration)}
\date{}
\title {\Large\textbf{Search for lepton-flavor-violating $\tau^- \to \ell^-\phi$ decays in 2019-2021 Belle II data}}

{\let\newpage\relax\maketitle}
    
\begin{abstract}
We report a search for lepton-flavor-violating decays $\tau^- \to \ell^- \phi$ ($\ell^- =e^-,\mu^-$) at the Belle II experiment, using a sample of electron-positron data produced at the SuperKEKB collider in 2019-2021 and corresponding to an integrated luminosity of 190\invfem.
We use a new untagged selection for $e^+e^- \to \tau^+\tau^-$ events, where the signal $\tau$ is searched for as a neutrinoless final state of a single charged lepton and a $\phi$ meson and the other $\tau$ is not reconstructed in any specific decay mode, in contrast to previous measurements by the BaBar and Belle experiments. 
We find no evidence for $\tau^- \to \ell^- \phi$ decays and obtain upper limits on the branching fractions at 90\% confidence level of 23 $\times 10^{-8}$ and 9.7$\times 10^{-8}$ for \ephi{} and \mphi{}, respectively.
\end{abstract}

\section{Introduction}
\label{sec:intro}

In the Standard Model (SM), lepton flavor is only accidentally conserved. The observation of neutrino oscillations not only introduces direct violation of neutrino flavor, but also charged-lepton flavor violation (LFV) through weak interaction charged currents. Decays that involve LFV are predicted to occur at rates close to $10^{-50}$\cite{ref:SM_cLFV, ref:tau_lfv}, far beyond the reach of current and future experiments. The observation of LFV decays would therefore provide indisputable evidence of non-SM physics.

Over the past four decades, CLEO II, experiments at the $B$ factories, and at the LHC have searched for LFV decays of the tau lepton, $\tau$. The most stringent upper limits on the branching fractions of several hypothetical tau decays are of the order of 10$^{-8}$~\cite{ref:HFL}. Among these are limits on the decays to final states with a lepton, either electron or muon, and a vector meson. In particular, in recent years there has been great interest in $\tau^- \to \ell^- \phi$ (charge-conjugated decays are implied throughout the paper) due to the $B$ meson anomalies reported by LHCb~\cite{ref:lhcb, ref:lfu_tests} and $B$-factory experiments~\cite{ref:HFL, ref:RD_babar, ref:RD_belle}. These measurements hint at potential lepton-flavor-universality violation in $b \to s \ell^+\ell^-$ and $b \to c \ell\bar{\nu}$ transitions, which might indicate a special role of the third lepton generation, with large enhancements of tau LFV decays. Specifically, leptoquark models that could accommodate both anomalies predict branching fractions for $\tau^- \to \mu^- \phi$ in the range from 10$^{-11}$ to a few times 10$^{-8}$~\cite{Angelescu:2021lln, Cornella:2021sby}, potentially accessible by the Belle II experiment with a data set of several inverse attobarns. 
The limits previously obtained by the BaBar and Belle experiments, using respectively 451\invfem\ and 854 fb$^{-1}$ of data, are given in Table \ref{table:prevUL}~\cite{BaBar:2009wtb, Belle:2011ogy}. Both experiments relied on the reconstruction of $e^+e^- \to \tau^+\tau^-$ events where one tau decays as \lphi\ and the other tau decays into a final state with a single charged particle (1-prong tagging).

Here, we report on a search for the \lphi\ decays based on an untagged selection, in which, in contrast to the previous searches, only the signal tau decay mode is explicitly reconstructed. Background events are suppressed by a set of requirements on topological and kinematic variables followed by a selection based on a boosted decision tree (BDT) algorithm that also includes kinematic variables and quantities related to the tracks and clusters not associated with the signal tau decays (rest-of-event variables). The signal is searched for in a two-dimensional plane $(M_{\tau}, \Delta E_{\tau})$, with $\Delta E_{\tau} = E^*_{\tau}-\sqrt{s}/2$ where $E^*_{\tau}$ and $M_{\tau}$ are respectively the reconstructed energy in the center-of-mass (c.m.)~frame and the invariant mass of the signal $\tau$ candidate, and $\sqrt{s}$ is the collision energy.
In cases where no significant signal is observed, upper limits are derived using the CLs method~\cite{Junk:1999kv, Read:2002hq}.
\begin{table}[!htb]
\caption{Expected and observed upper limits at 90\% C.L. on \lphi{} branching fractions obtained by BaBar~\cite{BaBar:2009wtb} (451 fb$^{-1}$) and Belle~\cite{Belle:2011ogy} (854 fb$^{-1}$). }
\begin{center}
\begin{tabular}{c|cc}
    \hline \hline
    \multirow{2}{*}{~~Experiment~~} & ~~$\mathcal{B}^{90}_{\mathrm{UL}}(e\phi)\ (\times 10^{-8})$~~ & ~~$\mathcal{B}^{90}_{\mathrm{UL}}(\mu\phi)\ (\times 10^{-8})$~~ \\
     & ~~exp. / obs.~~ & ~~exp. / obs.~~ \\
    \hline
    ~~BaBar~~ & ~~$5.0$ / $3.1$~~ & ~~$8.2$ / $19$~~ \\
    ~~Belle~~ & ~~$4.3$ / $3.1$~~ & ~~$4.9$ / $8.4$~~ \\
	\hline \hline
\end{tabular} 
\label{table:prevUL}
\end{center}
\end{table}

\section{The Belle II detector and Data Set}

Belle II~\cite{ B2TDR} is a particle physics experiment operating at the SuperKEKB asymmetric-energy electron-positron collider, located at KEK in Tsukuba, Japan. The energies of the
electron and positron beams are respectively 7 GeV and 4 GeV with a boost of the c.m. frame $\beta\gamma$ = 0.28 relative to the laboratory frame. The detector is composed of multiple layers of subdetectors arranged in a cylindrical structure with a central barrel and forward and backward endcaps. The $z$ axis of the laboratory frame is defined as the symmetry axis of the solenoid, pointing approximately along the incoming electron beam. The innermost component is the vertex detector, consisting of two inner layers of
silicon pixels and four outer layers of silicon strips. The
second pixel layer is partially installed, covering one sixth
of the azimuthal angle. The main tracking subdetector
is a large helium-based small-cell drift chamber (CDC) that also provides measurements of ionization-energy losses used for charged-particle identification. 
Outside the drift chamber, a Cherenkov-light imaging and time-of-propagation detector provides charged-particle identification in the barrel region, complemented in the forward endcap by a proximity-focusing, ring-imaging Cherenkov detector with an aerogel radiator. An electromagnetic calorimeter (ECL) consists of a barrel and two endcap sections made of CsI(Tl) crystals. A uniform 1.5 T magnetic field aligned with the beams is provided by a superconducting solenoid situated outside the calorimeter. Multiple layers of scintillators and resistive-plate chambers, located between the magnetic flux-return iron plates, constitute
the $K^0_L$ and muon identification system (KLM). 

The data used in this analysis are collected from electron-positron collisions at a c.m. energy $\sqrt{s} = 10.58$ GeV between March 2019 and July 2021 and correspond to an integrated luminosity of 190\invfem.

In order to define a selection to suppress backgrounds and estimate the signal efficiencies, we use simulated samples. For the study of the signal processes, we use 2 million  $e^+e^- \to \tau^+\tau^-$ events for each \lphi\ mode, where one tau decays into the LFV channel and the other into a known final state. In addition, we use simulated background samples consisting of $e^+e^- \to \tau^+\tau^-$ events generated with KKMC~\cite{Jadach_2000} in which tau decays are handled by  TAUOLA~\cite{Jadach:1993hs} and their radiative corrections by PHOTOS~\cite{ref:photos}; $\Upsilon(4S) \to B\bar{B}$ generated with EvtGen~\cite{Lange:2001uf}; $e^+e^- \to q\bar{q}$ $(q=u,d,s,c)$ continuum events generated with KKMC and quarks hadronization simulated with PYTHIA~\cite{Sj_strand_2015}; $e^+e^- \to \mu^+\mu^-(\gamma)$ generated with KKMC and $e^+e^- \to e^+e^-(\gamma)$ Bhabha-scattering background generated with BABAYAGA.NLO~\cite{Carloni_Calame_2000,Balossini_2008}. The total generated luminosity corresponds to 6\invapto\ for tau pairs and continuum samples, 1\invapto\ for the $ B\bar{B}$ and dimuon final states and 100\invfem\ for Bhabha events. 

The detector response is simulated with Geant4~\cite{GEANT4:2002zbu}. Data sets are analyzed using the Belle II analysis
software \textit{basf2}~\cite{Kuhr_2018, ref:zenodo}, which also provides a simulation of the trigger system.
 
\section{Candidate reconstruction and event selection}
\label{sec:reco}
This search exploits an online event selection (trigger) mainly relying on information provided by the ECL. Events are required to satisfy a minimal set of requirements on the energy depositions in the ECL (clusters) and their minimum energies. The most relevant is the requirement to have at least three ECL clusters, of which at least one is above 300 MeV.  Moreover, events consistent with Bhabha scattering are rejected by a dedicated trigger veto based on the energies and angles of two clusters in the ECL.

The $\tau$ lepton pairs produced in \emep\ annihilations are back-to-back in the c.m.~frame. Their decay products are well separated into two opposite hemispheres, defined by the plane perpendicular to the thrust axis $\mathbf{\hat{n}}_T$, which is the vector maximizing the quantity
\begin{equation}
\label{eq:thrust}
	T = \max_{\mathbf{\hat{n}}_T} \left(\dfrac{\sum_{i} \left|\mathbf{p}_i \cdot \mathbf{\hat{n}}_T\right|}{\sum_{i} \left|\mathbf{p}_i\right|} \right),
\end{equation}
where $\mathbf{p}_i$ is the momentum of final state particle $i$, including both charged and neutral particles.

In the offline reconstruction, charged-particle trajectories (tracks) are required to come from the interaction point (IP), with transverse ($dr$) and longitudinal ($dz$) projections of the distance of the closest approach respectively less than 1 and 3 cm in absolute value.

Identification of muons relies mostly on charged-particle penetration depth in the KLM for momenta larger than 0.7 GeV/$c$ and on information from the CDC and ECL otherwise. Particle-identification likelihoods are obtained by combining information from the relevant subdetectors.
The ratio between the muon likelihood and the sum of the likelihoods of all particle hypotheses is required to exceed 0.99, resulting in 86.9\% efficiency for selecting muon candidates with momenta in the range $ 1.0 < p < 1.5$~\gevc and within the KLM barrel acceptance. The rate for a pion to be misidentified as a muon is 2.8\%. 

\begin{figure}[h!]
	\makebox[\textwidth][c]{
		\begin{minipage}[t]{0.5\textwidth}
			\centering
			\includegraphics[width=1\textwidth]{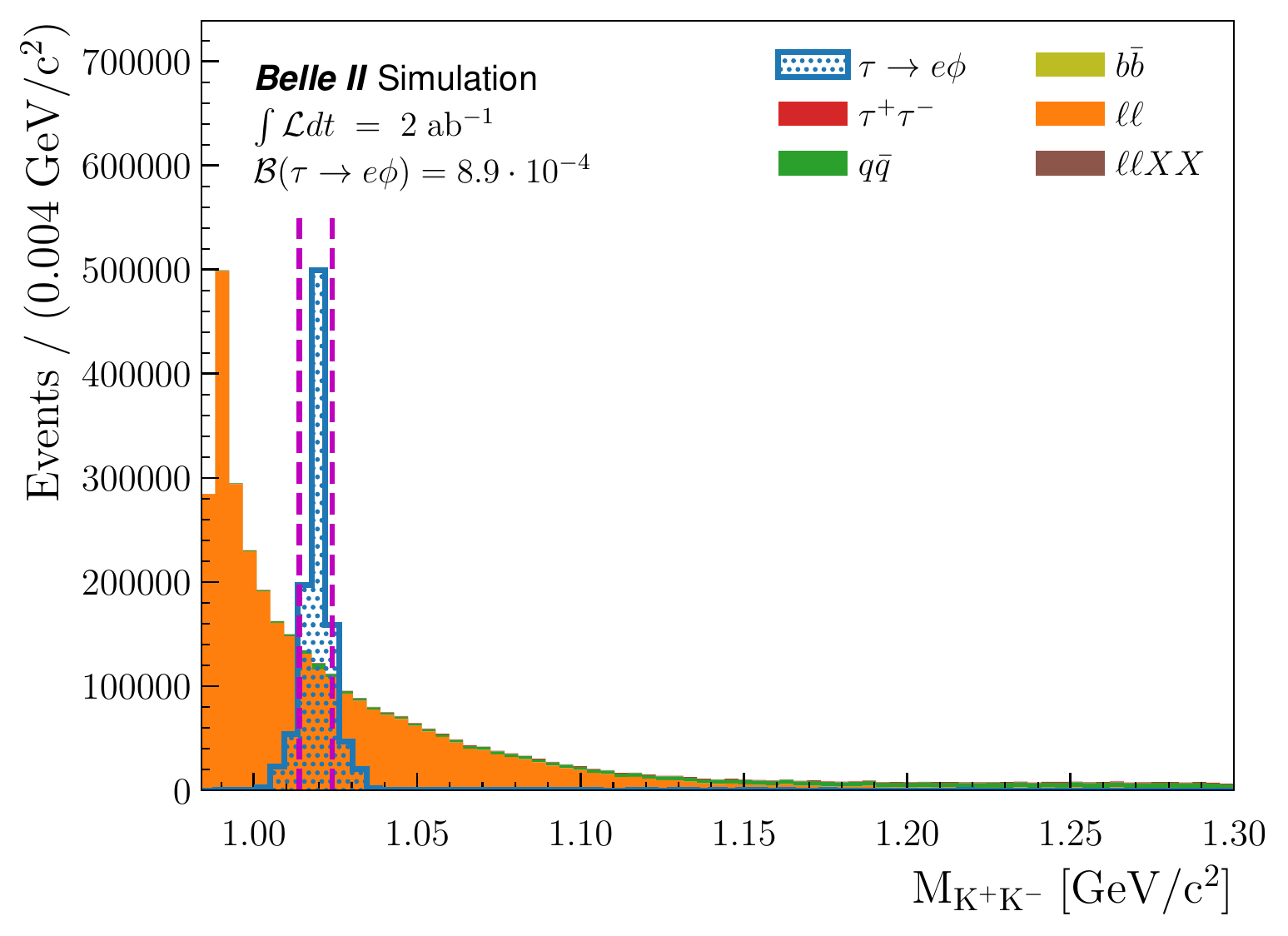}
		\end{minipage}%
		\hspace{0.2cm}%
		\begin{minipage}[t]{0.5\textwidth}
			\centering
			\includegraphics[width=1\textwidth]{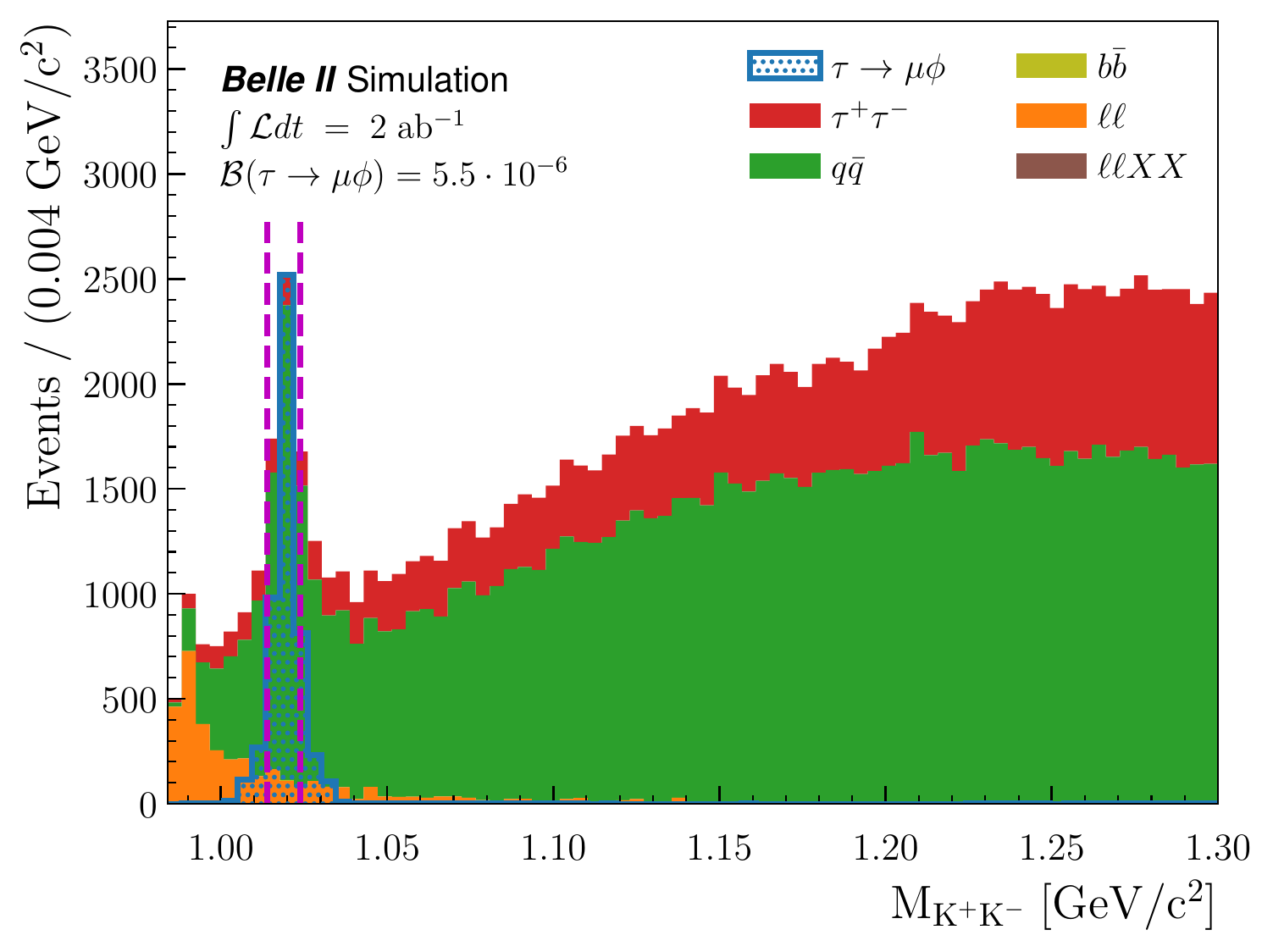}
		\end{minipage}
	}
	\caption[]{Reconstructed $\phi$ candidate invariant mass $M_{K^+K^-}$ for the electron (left) and muon (right) modes after the candidate reconstruction and selections up to the $\phi$ candidate mass window definition. The solid stacked histograms are the simulated background samples for an equivalent generated luminosity of 2\invapto, while the blue dotted histogram represents the signal simulation for $\tau\to \ell \phi$ decays, scaled to arbitrary branching fractions ($\mathcal{B}$). The vertical dashed lines show the edges of the region where events are retained for further analysis.}
	\label{fig:phi_mass}
\end{figure}

Electrons are primarily identified by comparing momenta measured by tracking with energies of associated ECL depositions. This information and additional measurements characterizing the calorimeter clusters are used as input to a boosted decision tree (BDT) classifier that combines ECL-based measurements with the other subdetectors’ normalized likelihoods. For charged-particle momenta in the range $ 1.0 < p < 1.5$~\gevc, the rate for pions to be misidentified as electrons is 0.1\% at 90\% selection efficiency for electron candidates. Additionally, electrons are corrected for bremsstrahlung losses by summing the four-momenta of any photon found within a 0.15 radian wide cone around the direction of the electron momentum and with a minimum energy of 20 MeV. Kaons are selected as tracks coming from the IP consistent with the kaon mass  hypothesis. For the kaon with the highest transverse momentum, we require that the ratio between the kaon likelihood and the sum of the likelihoods of all particle hypotheses is greater than 0.5. A $\phi$ candidate is then reconstructed as a pair of oppositely charged kaons with an invariant mass $1.014 < M_{K^+K^-} < 1.024$ GeV/$c^2$. The selected mass region, indicated by the vertical dashed red lines in Figure~\ref{fig:phi_mass}, is optimized by maximizing the Punzi figure of merit~\cite{Punzi:2003bu}, configured for a search sensitivity at 3$\sigma$ level (in units of one-sided Gaussian).

The signal $\tau$ candidate is reconstructed as one lepton candidate, either \textit{electron} or \textit{muon}, which defines the search channel mode, 
and one $\phi$ candidate belonging to the same hemisphere. 
The total number of reconstructed tracks in the event is required to be smaller or equal to six. No attempt is made to explicitly reconstruct the other $\tau$ lepton. 
Instead we use the inclusive properties of all other particles in the event to suppress backgrounds.

Photons are reconstructed from ECL clusters within the CDC acceptance and not associated with any track. Photons used for $\pi^0$ reconstruction are  required to have an energy deposit of at least 0.1 GeV. Neutral pions are then identified as photon pairs with masses within $0.115 < M_{\gamma\gamma}< 0.152$ GeV/$c^2$. Photons used to reconstruct $\pi^0$ candidates  are not considered for electron bremsstrahlung correction. Photons that satisfy the above conditions, but are not used for bremsstrahlung correction, nor in the $\pi^0$ reconstruction, and with energies greater than 0.2 GeV are used to define the thrust and the CLEO cones~\cite{ref:cleoCone}. The latter are variables computed from the momentum flow of the reconstructed particles binned in 10$^{\circ}$ cones around a given axis (e.g. the collision or thrust axis)~\cite{ref:physicsAtBfac}.

All the tracks and clusters that are not used in the signal reconstruction form the rest of the event (ROE), whose kinematic properties are exploited for further background suppression. The following criteria must be satisfied to be retained as part of the ROE: photons are required to have energies of at least 50 MeV and tracks must have a transverse momentum higher than 50 MeV$/c$ with $|dz| < 10\ \mathrm{cm}$ and $|dr| < 5\ \mathrm{cm}$.

We define rectangular regions in the two-dimensional plane ($M_\tau, \Delta E_\tau$). The distribution of \deltae\ is expected to peak at zero for correctly reconstructed signal candidates and \invmass\ peaks at the known $\tau$ mass~\cite{ParticleDataGroup:2022pth}. The rectangular boxes are centered at the signal peak in the ($M_\tau, \Delta E_\tau$) plane, with side lengths equal to multiples of their resolutions $\delta$. For each variable, $\delta$ is estimated by fitting the sum of a Crystal Ball and two Gaussian functions to the signal distribution, and then taking the weighted average of the three respective widths.
Their values are reported in Table~\ref{tab:sig_resol}.

\begin{figure}[h!]
  \makebox[\textwidth][c]{
    \begin{minipage}[t]{0.5\textwidth}
      \centering
      \includegraphics[width=1\textwidth]{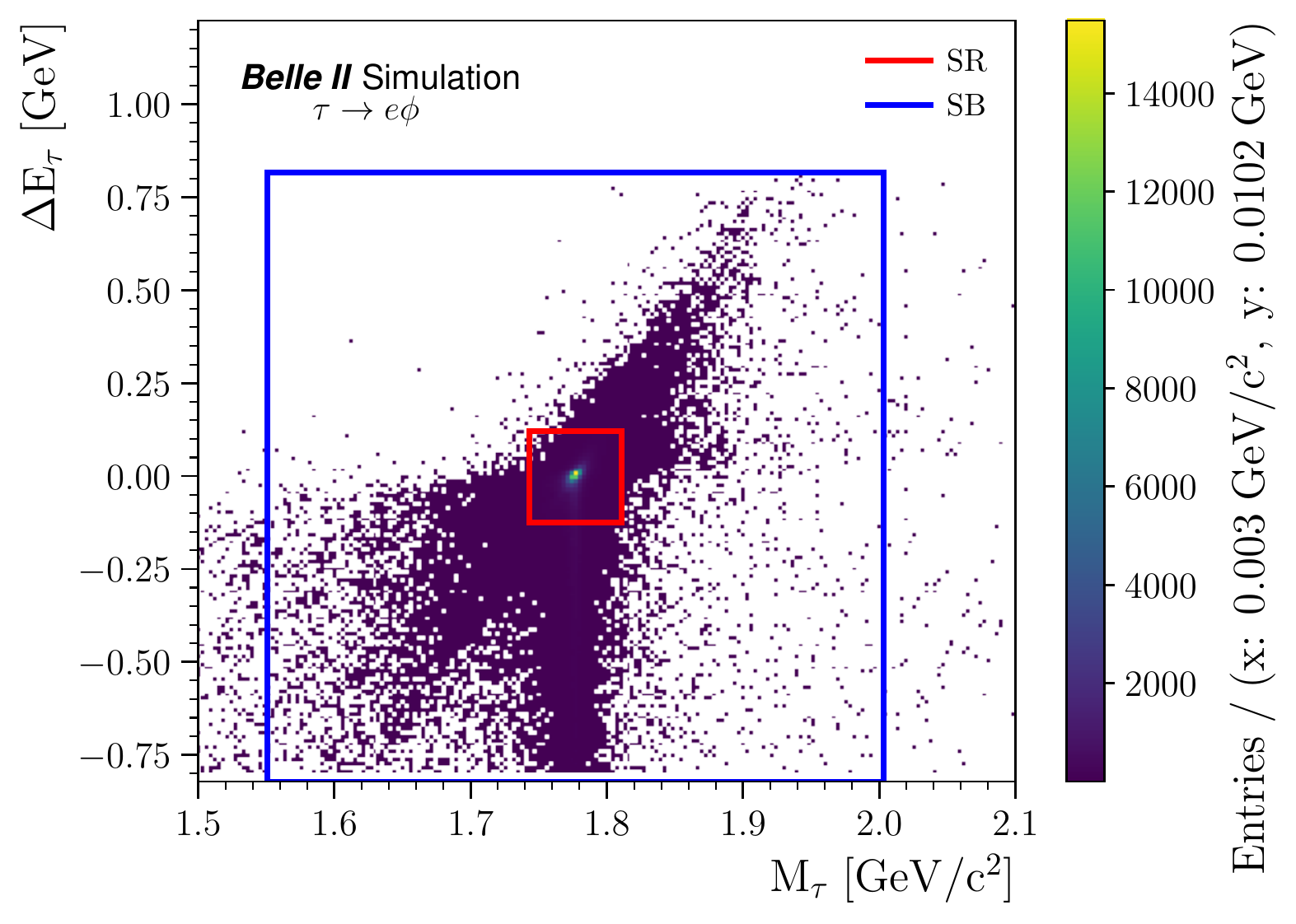}
    \end{minipage}%
    \hspace{0.2cm}%
    \begin{minipage}[t]{0.5\textwidth}
      \centering
      \includegraphics[width=1\textwidth]{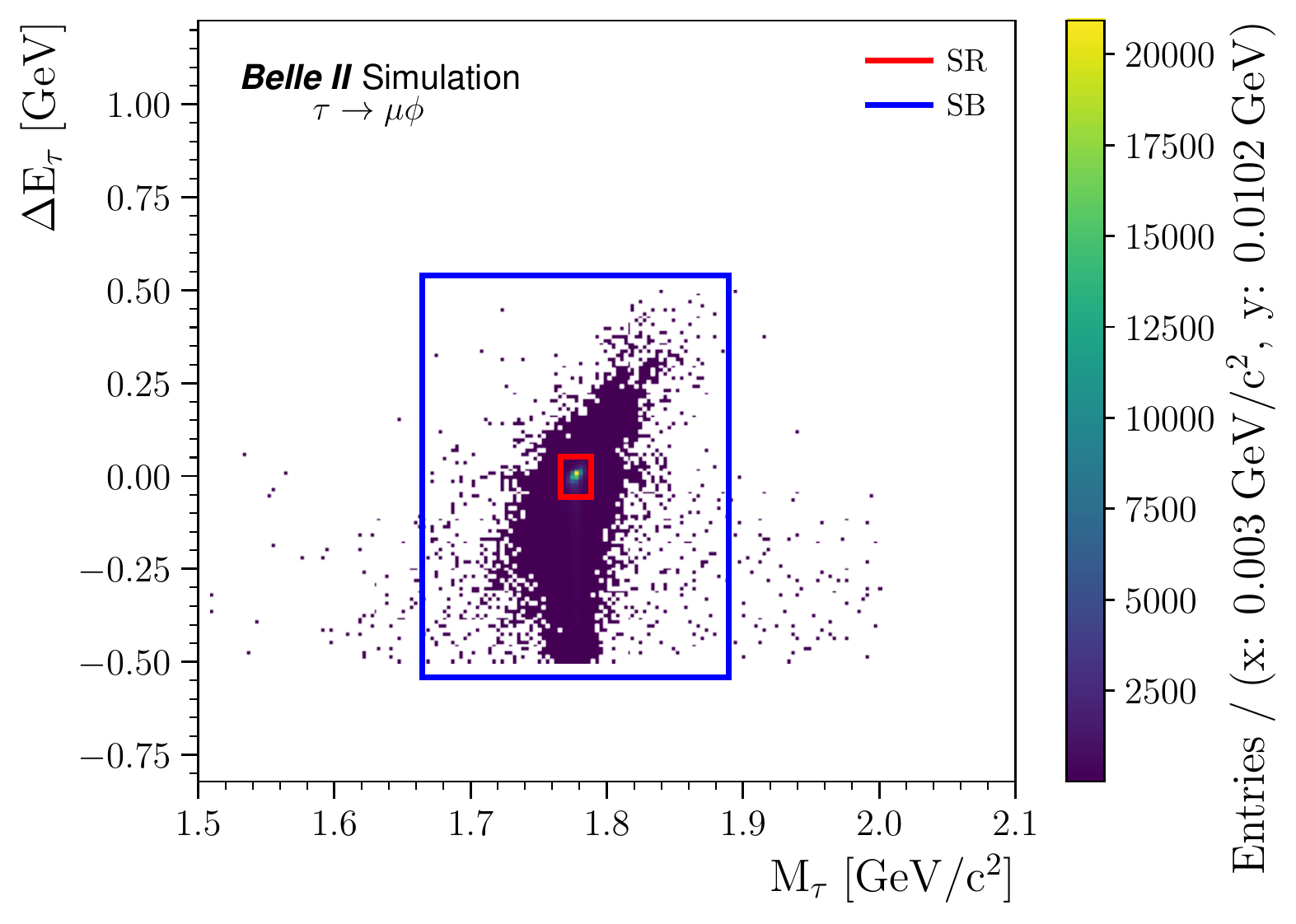}
    \end{minipage}
  }
\caption{Reconstructed events for the simulated signal samples in the $M_{\tau}$-$\Delta E_{\tau}$ plane. The edges of the $\pm 3\delta$ for $\tau \to e\phi$, $\pm 2\delta$ for $\tau \to \mu\phi$, and $\pm 20\delta$ regions are marked as red and blue rectangles, respectively, for the electron (muon) mode in the left (right) plot. The sidebands (SB) are the areas inside the blue boxes and outside the red ones.  }
\label{fig:signalRegion}
\end{figure}

\begin{table}[!htb]
\begin{center}
    \caption{Resolutions $\delta$ obtained from the fits to \invmass\ and \deltae\ distributions on the simulated electron (muon) signal sample.}
\begin{tabular}{c|cc}
    \hline \hline
	~~ Mode~~ & ~~ Variable ~~ & ~~$\delta$~~ \\
\hline
	 \multirow{2}{*}{~~$e\phi$~~} & ~~$M_{\tau}$~~ & ~~$11.3 \pm 0.2\ \mathrm{MeV}/c^2$~~ \\
	  & ~~$\Delta E_\tau$~~  & ~~$41.0 \pm 0.8\ \mathrm{MeV}$~~ \\
	  \hline
	 \multirow{2}{*}{~~$\mu\phi$~~} & ~~$M_{\tau}$~~ & ~~$5.6 \pm 0.2\ \mathrm{MeV}/c^2$~~ \\
	  & ~~$\Delta E_\tau$~~ & ~~$27.1 \pm 0.6\ \mathrm{MeV}$~~ \\
	\hline \hline
\end{tabular}
\label{tab:sig_resol}
\end{center}
\end{table}
A $\pm 20\delta$ wide rectangular region is used for the selection optimization on simulation, while all reconstructed events outside this box in the (\invmass, \deltae) plane are rejected. The $\pm3\delta$ and $\pm 2\delta$ boxes are used as signal regions (SR) for the final signal-yield extraction in the electron and muon modes, respectively. Events inside the above-defined signal regions in data are hidden during the optimization of the selection to avoid introducing experimental bias. These regions are illustrated in \figurename~\ref{fig:signalRegion}.  Structures due to the initial and final state radiation, seen respectively as longer tails in the negative \deltae\ and \invmass\ directions, are visible in the plots. Final state radiation affects mainly the electron mode and it is corrected for in the event reconstruction by taking into account the bremsstrahlung losses.

\section{Background suppression}
\label{sec:BDT}
The background originates from radiative dilepton and four-lepton final states (\textit{low-multiplicity}), incorrectly reconstructed SM \emep$\rightarrow\tau^+\tau^-$ and continuum hadronization processes from \emep$\rightarrow q\bar{q}$ events.
To remove the low-multiplicity background, we exploit the zeroth order CLEO cone; the angle between the momentum of the reconstructed signal $\tau$ and the closest track ($\theta_{\tau-\mathrm{closest}}$); and the opening angle between the two reconstructed kaons ($\theta_{K1-K2}$) for the electron mode. We require the zeroth CLEO cone to be less than $8.5$ GeV/$c$; $\theta_{\tau-\mathrm{closest}}$ larger than 0.02 rad, and $\theta_{K1-K2}$ less than 0.01~rad, respectively. We select events with thrust axis magnitude $T<0.99$ for the muon mode. The reconstruction and selection described above reject 97.9\% of the total simulated background for the electron channel and 86.3\% for the muon channel, and retain 11\% and 9\% of signal candidates inside the $\pm 3\delta$ and the $\pm 2 \delta$ signal regions for the electron and muon modes, respectively.

The remaining background events are mainly from continuum processes and are further suppressed by combining event-shape variables, kinematic properties of the signal $\tau$ candidate, of the reconstructed $\phi$ resonance, of the final-state particles, and variables related to the ROE, into a binary BDT classifier implemented with the XGBoost library~\cite{XGBoostPaper}. The BDT is trained over a portion of the simulated signal and background events. 
The optimal selection on the BDT output is determined by minimizing the upper limit computed from the expected number of background events in the simulation in the $\pm 5\delta$ wide boxes in the (\invmass, \deltae) plane. 

After the application of the BDT selection, there is a significant contribution to the remaining background observed in data control regions for the electron mode from the $V^0$-photoproduction process \emep$\rightarrow e^+e^-\phi$. 
No simulation of this process is available. Therefore we devise a data driven veto to reject this contribution: events reconstructed in the electron mode are rejected if they have more than one electron candidate in the final state or the second-order CLEO cone with respect to the collision axis is larger than 0.4 or the cosine of the angle between the reconstructed signal $\tau$ momentum and its direction of flight is smaller than zero.

The signal efficiency, extracted from the remaining independent sample as the ratio between events passing the selections inside the signal region and the total generated events, is corrected for the measured discrepancies in the particle-identification efficiency between data and simulation. 
The final efficiencies for the electron and muon modes are respectively $(6.1 \pm 0.9)\%$ and $(6.5 \pm 0.6)\%$, where the quoted uncertainty is the total systematic uncertainty computed from the list in Table~\ref{tab:syst}. 

\begin{figure}[h!]
  \makebox[\textwidth][c]{
    \begin{minipage}[t]{0.5\textwidth}
      \centering
      \includegraphics[width=1\textwidth, page=1]{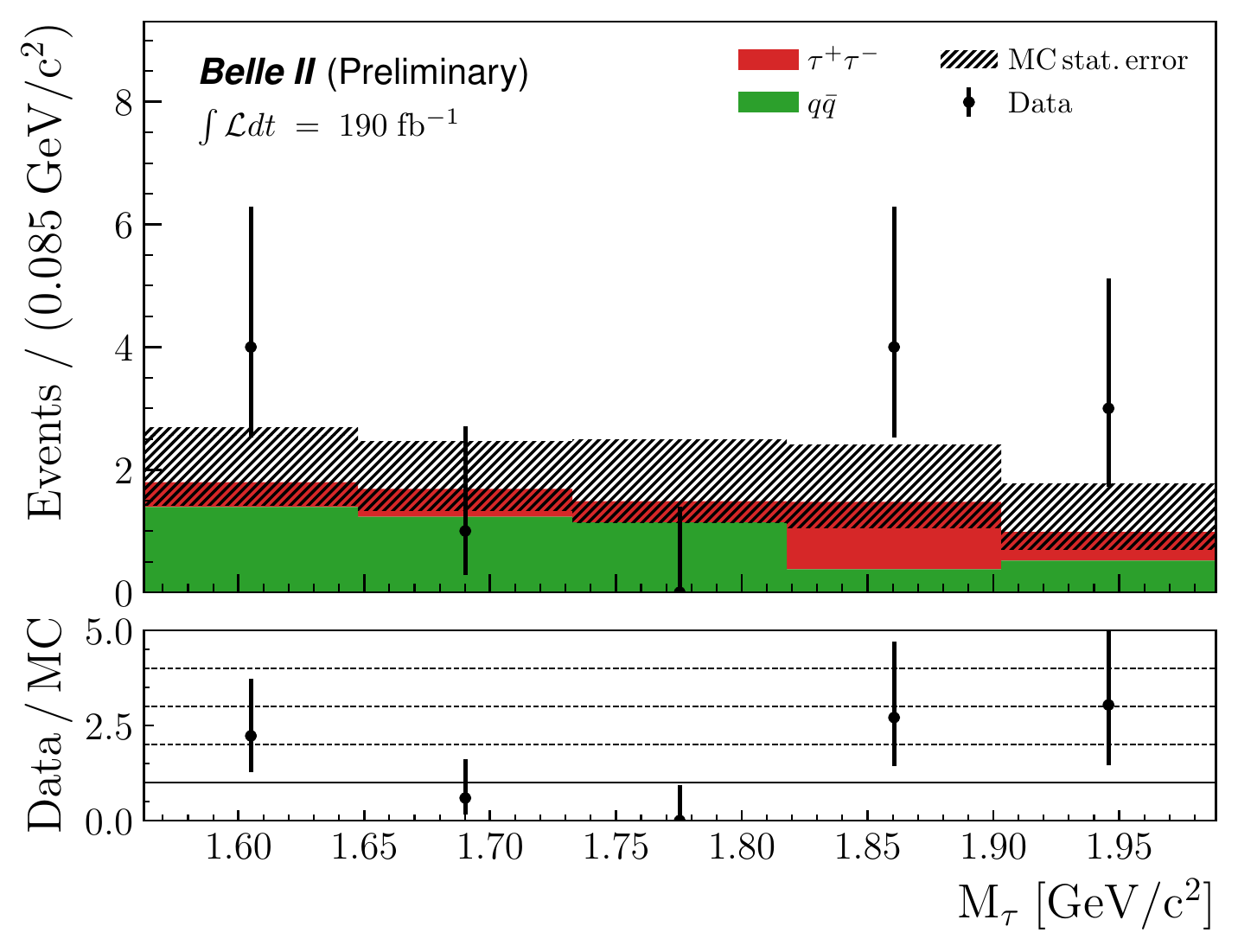}
    \end{minipage}%
    \hspace{0.2cm}%
    \begin{minipage}[t]{0.5\textwidth}
      \centering
      \includegraphics[width=1\textwidth,page=2]{plots/elphi_DataMC_BDT_dataCut_mini_SB.pdf}
    \end{minipage}
  }
  \makebox[\textwidth][c]{
    \begin{minipage}[t]{0.5\textwidth}
      \centering
      \includegraphics[width=1\textwidth, page=1]{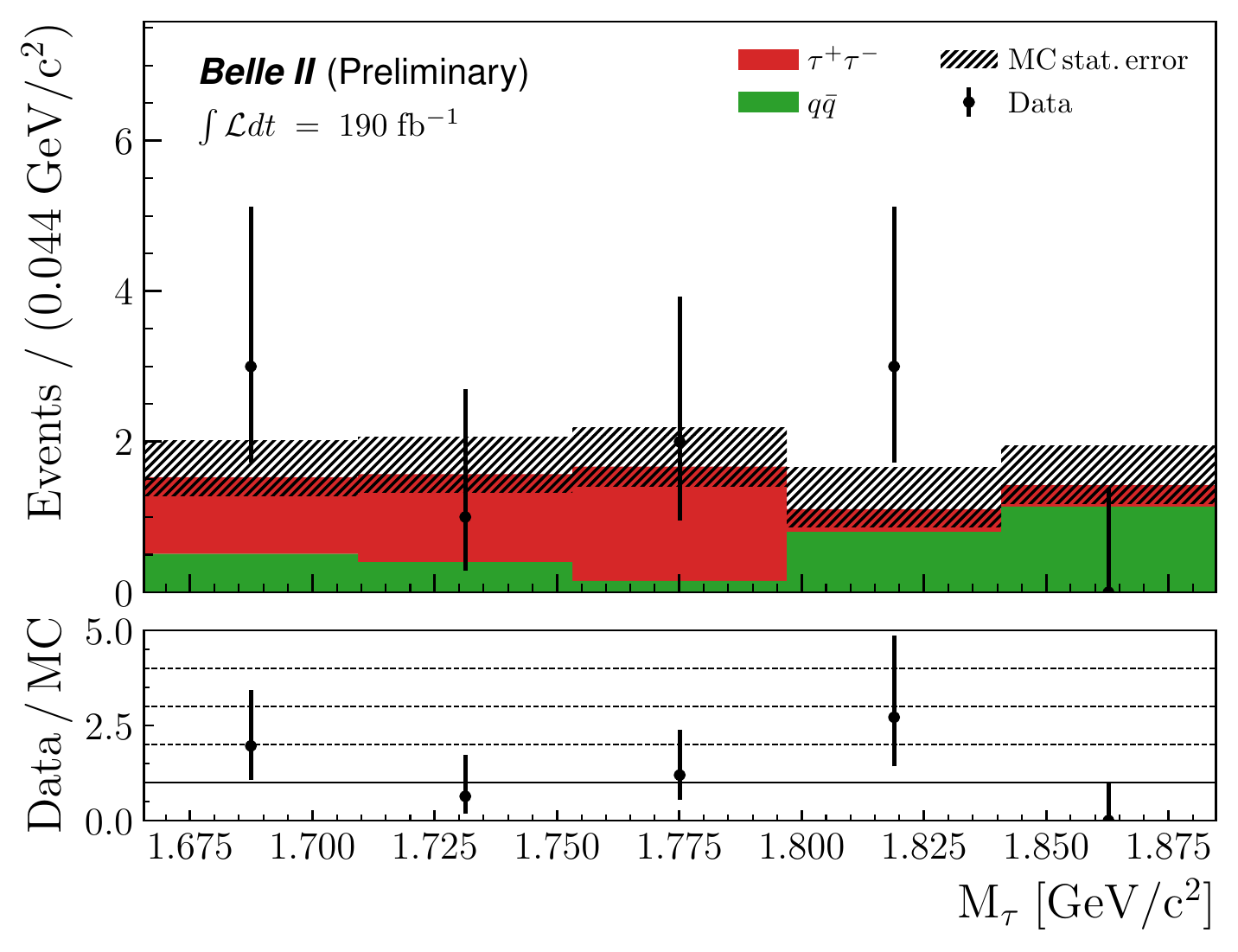}
    \end{minipage}%
    \hspace{0.2cm}%
    \begin{minipage}[t]{0.5\textwidth}
      \centering
      \includegraphics[width=1\textwidth,page=2]{plots/muphi_DataMC_BDT_dataCut_mini_SB.pdf}
    \end{minipage}
  }
\caption{ Comparison of the $M_\tau$ (left) and $\Delta E_\tau$ (right) distributions in the sideband regions of data (the points with error bars) and simulation (colored stacked histograms), for the \ephi{} (top) and \mphi{} (bottom) channels, after all selection requirements. Subpanel plots show the data to MC ratios, consistent with unity within the uncertainties.}
\label{fig:DataMCBDT}
\end{figure}

The agreement between data and simulation after the final selections is validated by using the region outside the signal boxes in the (\invmass, \deltae) plane, from $\pm 3\delta$ for the electron and from $\pm 2\delta$ for the muon mode up to $\pm 20\delta$, corresponding to the area outside the red box and inside the blue one in the plots in Figure~\ref{fig:signalRegion}. The $M_\tau$ and $\Delta E_\tau$ distributions in the sideband regions after the final selections on data and simulation are shown in Figure~\ref{fig:DataMCBDT}. The simulated samples are scaled to data luminosity and corrected for the differences observed between data and simulation in particle identification efficiencies. 
The numbers of expected background events from the simulation are 7.4$^{+1.1}_{-1.0}$ and 7.3$^{+0.8}_{-0.7}$ for the electron and muon modes, where the associated asymmetric errors are statistical; in data, we observe 12.0$^{+4.5}_{-3.4}$ and 9.0$^{+4.1}_{-2.9}$ events, respectively. This results in  ratios between data and simulation of, respectively, 1.6$^{+0.6}_{-0.5}$ and 1.2$^{+0.5}_{-0.4}$, which shows a data and simulation agreement compatible with unity. 

\section{Yield extraction}
\label{sec:yields}

We estimate the final yields in a Poisson-experiment approach by counting the number of observed events in data inside the SR. To estimate the number of expected background events $N_\mathrm{exp}$ inside the SR, we use the \invmass\ sidebands within the $\pm 3(2)\delta$ \deltae\ band, defined as reduced sidebands (RSB). Therefore, the expected background is
\begin{equation}
\label{eq:bkgEstimate}
    N_\mathrm{exp}= N_\mathrm{data} \times r_\mathrm{MC}
\end{equation}
where $N_\mathrm{data}$ is the yield in the data inside the RSB regions and $r_{MC}$ is the ratio measured on simulation between the events inside the signal box and those retained in the RSB region.

\begin{figure}[h!]
  \makebox[\textwidth][c]{
    \begin{minipage}[t]{0.5\textwidth}
      \centering
      \includegraphics[width=1\textwidth]{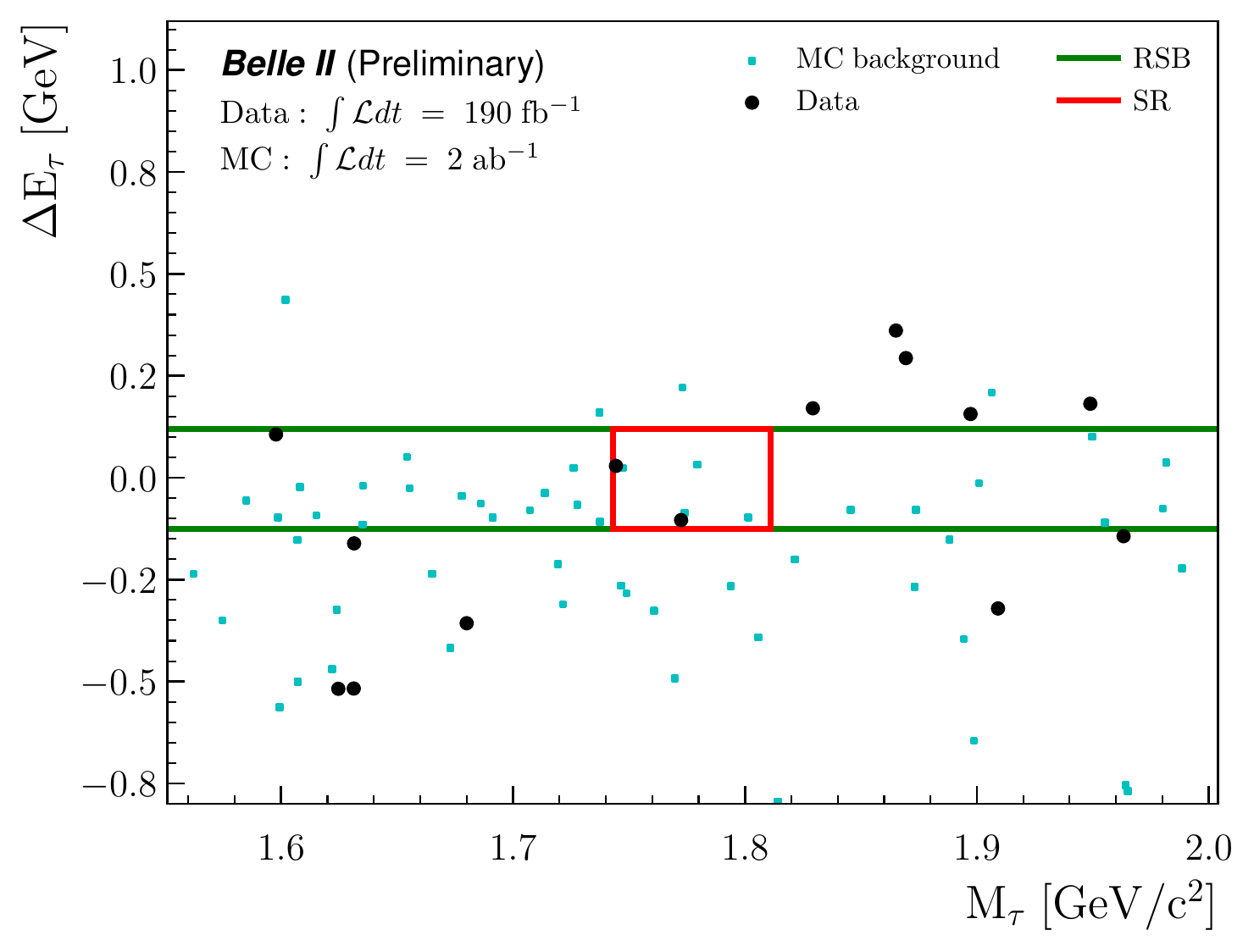}
    \end{minipage}%
    \hspace{0.2cm}%
    \begin{minipage}[t]{0.5\textwidth}
      \centering
      \includegraphics[width=1\textwidth]{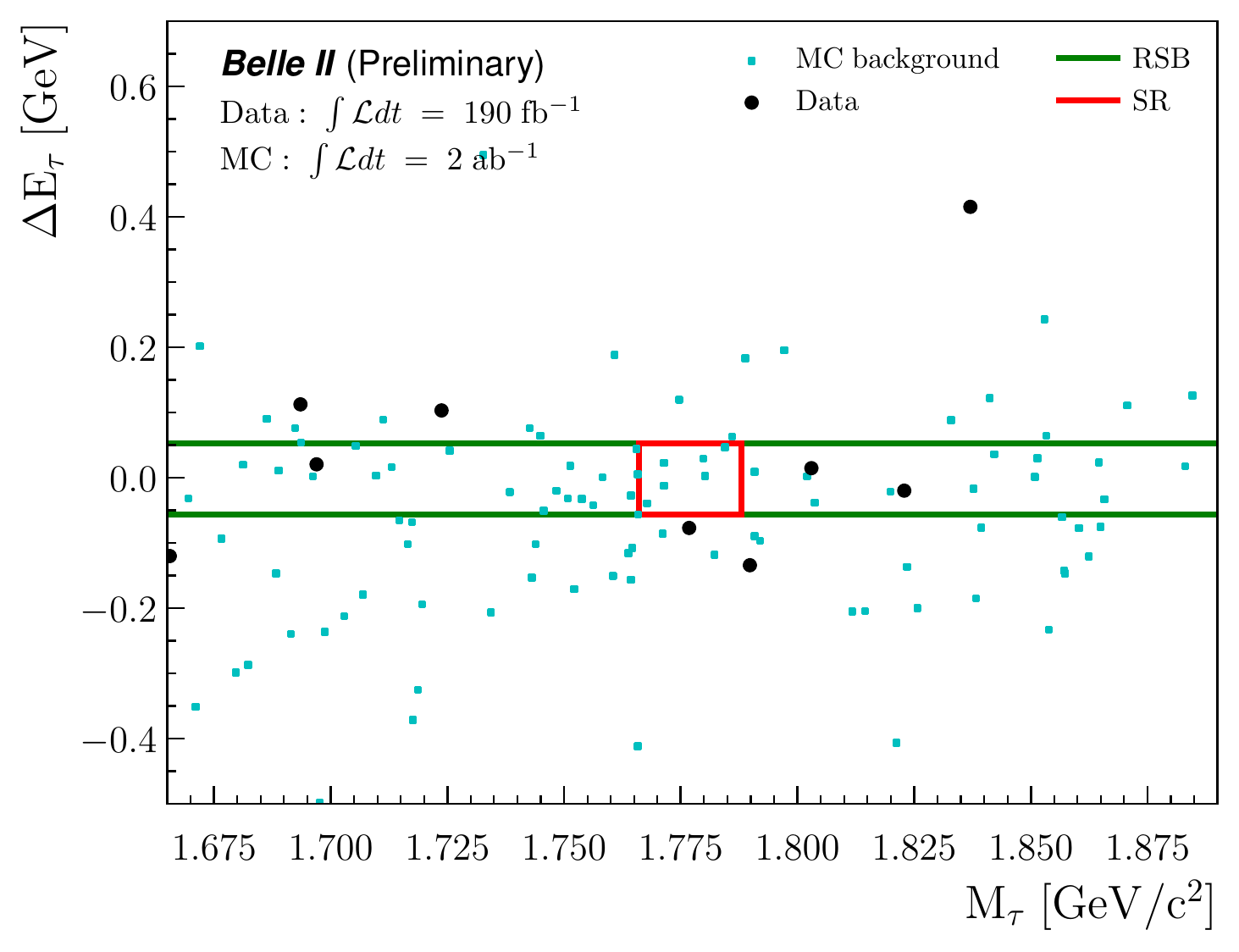}
    \end{minipage}
  }
\caption[Scatter plots of  $\Delta E_\tau$ vs $M_\tau$ for simulated SM background and  data in the \ephi{} and \mphi{} channels, after all selections.]{Scatter plots of 
 $\Delta E_\tau$ vs $M_\tau$ for simulated SM background and  data in the \ephi{} (left) and \mphi{} (right) channels, after all selections. The red squares represent the signal regions (SR), while the green lines delineate the $\pm 3\delta$ ($e\phi$) or $\pm 2\delta$ ($\mu\phi$) $\Delta E_\tau$ band (RSB) used for estimating the expected number of events in the SR. The RSB region is the area outside the red box and inside the green solid lines. }
\label{fig:ScPlotsData}
\end{figure}

The MC ratios are respectively $0.23^{+0.16}_{-0.10}$ and $0.12^{+0.07}_{-0.04}$ for the electron and muon channels, where the associated uncertainties are statistical only, since we assume that any possible systematic contribution due to the simulation, reconstruction, or selections cancel out in the computed ratio. Having observed one and three events in the data RSB for the electron and muon modes, respectively, we compute the expected background in the signal boxes to be 0.23 for the electron mode and 0.36 for the muon mode. The results are summarized in Table~\ref{tab:finalRes}. 
Figure~\ref{fig:ScPlotsData} shows the remaining events in the (\invmass, \deltae) plane for data and simulated samples after the final selections. 

\begin{table}[!htb]
\caption[]{Final selection efficiencies and yields on data and simulation for specific regions of the ($M_\tau$, $\Delta E_\tau$) plane: the signal regions (SR) and the reduced sidebands (RSB) for both channels. The expected data yield in the SR $N_{\mathrm{exp}}$ is the product of the yield $N_\mathrm{data}$ observed in the RSB and the MC background ratio $r_\mathrm{MC}$. The final observed yields in the data SR are also reported in the last row. Statistical (stat) and systematic (syst) uncertainties are omitted wherever negligible.}
\begin{adjustbox}{max width=1.3\textwidth, center}
\begin{tabular}{ccccc}
\hline \hline
\multicolumn{2}{c}{\multirow{2}{*}{~~Quantity~~}} & \multirow{2}{*}{~~Region~~} & \multicolumn{2}{c}{~~Mode~~} \\
 & & & ~~$e\phi$~~ & ~~$\mu\phi$~~ \\
\hline
\multicolumn{2}{c}{~~\multirowcell{2}[0pt][c]{Signal\\efficiency $\varepsilon_{\mathrm{\ell\phi}}$}~~} & \multirow{2}{*}{~~SR~~} & \multirow{2}{*}{~~(6.1\unskip\ $\pm$ 0.9\unskip$~\mathrm{(syst)}$)\%~~} & \multirow{2}{*}{~~(6.5\unskip\ $\pm$ 0.6\unskip$~\mathrm{(syst)}$)\%~~} \\
 &  &  & \\ 
\hline
\multicolumn{2}{c}{~~\multirowcell{2}[0pt][c]{~~$r_\mathrm{MC}$}~~} & \multirow{2}{*}{~~SR / RSB~~} & \multirow{2}{*}{~~$0.23 ^{+0.16}_{-0.10}$~$\mathrm{(stat)}$~~} & \multirow{2}{*}{~~$0.12^{+0.07}_{-0.04}$
~$\mathrm{(stat)}$~~} \\
 &  &  & \\
\hline
~~\multirowcell{2}[0pt]& ~~$N_{\mathrm{data}}$~~ 
 & ~~RSB~~ & ~~1.0$^{+2.3}_{-0.8}$~$\mathrm{(stat)}$~~ & ~~3.0$^{+2.9}_{-1.6}$~$\mathrm{(stat)}$~~ \\
 & ~~$N_{\mathrm{exp}}$~~ & ~~SR~~ & ~~0.23$^{+0.55}_{-0.21}$~$\mathrm{(stat)}$~~ & ~~0.36$^{+0.39}_{-0.23}$~
 $\mathrm{(stat)}$~~ \\
 & ~~$N_{\mathrm{obs}}$~~ & ~~SR~~ & ~~2.0$^{+2.6}_{-1.3}$~$\mathrm{(stat)}$~~ & ~~0.0$^{+1.8}_{-0.0}$~
 $\mathrm{(stat)}$~~ \\
\hline \hline
\end{tabular}
\end{adjustbox}
\label{tab:finalRes}
\end{table}

\section{Upper limit computation}
\label{sec:ul}
The 90\% confidence level (C.L.) upper limits on the \lphi{} branching fractions are computed as
\begin{equation}\label{eq:result}
 \mathcal{B}_{\mathrm{UL}}(\tau \to \ell\phi) = \frac{s}{L \times 2\sigma_{\tau\tau} \times \varepsilon_{\mathrm{\ell\phi}}},
\end{equation}
where $s$ is the corresponding 90\% C.L. upper limit on $N_{\mathrm{obs}} - N_{\mathrm{exp}}$, the difference between the observed and expected data yields. The other inputs to Equation~\ref{eq:result} are the integrated luminosity $L=190\pm 1~\mathrm{fb^{-1}}$ of the analyzed data sample; the $\tau$-pair production cross section $\sigma_{\tau\tau}= 0.919\pm 0.003~\mathrm{nb}$ and the signal efficiency $\varepsilon_{\mathrm{\ell\phi}}$. The selection of a specific final state for the $\phi$ candidate reconstructed as $\phi\rightarrow K^+K^-$ is already included in the total signal efficiency $\varepsilon_{\mathrm{\ell\phi}}$, as the simulated samples allow for any possible known decays of the $\phi$. Therefore no normalization to $\mathcal{B}(\phi\rightarrow K^+K^-)$ is needed in Equation~\ref{eq:result}.

We estimate upper limits with the CL$_s$~\cite{Junk:1999kv, Read:2002hq} method in a frequentist approach implemented in the \texttt{pyhf} library~\cite{pyhf, pyhf_joss}. We generate $10000$ toys at 50 points distributed uniformly in the branching fraction range $(0 - 0.5 \times 10 ^{-7})$. Results in Table~\ref{tab:finalRes} are taken as inputs to Equation~\ref{eq:result} and used to compute the branching fraction value at each scan point. The total statistical and systematic uncertainties affecting each of the experimental inputs are combined in quadrature. The total relative uncertainty on $L \times 2\sigma_{\tau\tau} \times \varepsilon_{\mathrm{\ell\phi}}$ is 15\% for the electron channel and 8.7\% for the muon, as discussed in Section~\ref{sec:syst}. 
The dominant uncertainty on $N_{\mathrm{exp}}$ is statistical and  is reported in Table~\ref{tab:finalRes} for each analyzed channel. Figure \ref{fig:CLs} shows the $\mathrm{CL}_s$ curves computed as a function of the upper limit on the branching fractions. The dashed black line shows the expected $\mathrm{CL}_s$ and the green and yellow bands give the $\pm1\sigma$ and 
$\pm2\sigma$ contours, respectively. The expected limits for the \ephi{} and \mphi{} modes are $15\times 10^{-8}$ and $9.9\times 10^{-8}$, respectively.
%
\section{Systematic uncertainties}
\label{sec:syst}
Systematic uncertainties arise from detector effects and differences between data and simulation, due to possible mismodeling in the generation and reconstruction of the simulated samples. A summary of considered systematic uncertainties on each parameter entering the measurement is provided in Table~\ref{tab:syst}.

Contributions to the systematic uncertainty on the total signal efficiency arise from uncertainties is particle-identification performance, trigger efficiency, track reconstruction efficiencies, and possible differences between data and simulation in signal distributions of variables used in the background-suppression selections.

Discrepancies in particle identification for leptons and charged hadrons between data and simulation are measured in control samples and the derived corrections are applied to the reconstructed events in simulation. The applied corrections are then varied within their statistical and systematic uncertainties, and the final selection efficiency for each configuration is calculated. We assign the maximum relative difference with respect to the nominal efficiency as the systematic contribution due to particle identification. This corresponds to a relative uncertainty of 0.8\% and 0.3\% for the electron  and the muon modes, respectively.

The difference between data and simulation in the track-reconstruction efficiency is measured  with tau-pair events 
and yields a 0.3\% systematic uncertainty per track. We assign a relative uncertainty of 0.9\% on the signal efficiency due to tracking.

The trigger efficiencies in data and simulation agree to better  than the 1\% level. We assign an uncertainty estimated from the signal simulation due to a possible bias in the method to compare trigger efficiencies between data and simulation. The absolute trigger efficiency, defined as the number of events in signal simulation that meet the ECL trigger selections required for this analysis divided by the total generated, cannot be measured directly in the data. Only the efficiency relative to orthogonal trigger lines with respect to those used for the signal selection can be accessed. The latter is calculated as the fraction of events passing the ECL trigger selections among all events selected by independent trigger selections based on CDC information. The differences of the ratios of these two efficiencies from unity are assigned as 0.4\% and 0.9\% systematic uncertainties for the electron and muon modes, respectively.

Potential differences in distributions between data and simulation of variables used in the selection may lead to a systematic uncertainty affecting the total signal efficiency. This is estimated from two control samples, \phipi{} and \threepi\ to ensure the maximum kinematic overlap with signal samples for the variables entering the BDT.
Variables showing the largest discrepancies between data and simulation are corrected bin-by-bin using the control samples.
We then retrain a BDT on the corrected samples and recompute the signal efficiency. We assess the systematic uncertainty as the difference with respect to the nominal efficiency values quoted in Table~\ref{tab:finalRes}. This procedure gives the largest contribution to the estimated systematic uncertainty, that  is 15.2\% and 8.5\% for the electron and muon modes, respectively.
To compensate for the imperfections of the magnetic-field map
 used in the event reconstruction, misalignment of the detector, and mismodeling in the density and type of detector and beam-pipe material,
we scale the charged-particle momenta by a correction factor, measured from the shift observed in data on the invariant mass of $D^0$ relative to its known value ~\cite{ParticleDataGroup:2022pth}. 
The corresponding systematic uncertainty is obtained
by varying the correction factor according to its  uncertainty, which leads to different data yields in the sideband region, and thus different numbers of expected events $N_\mathrm{exp}$. The resulting systematic uncertainty, taken as the difference with the nominal value, is 0.6\% (0.4\%) for the electron (muon) mode.
The systematic uncertainty on the integrated luminosity $L$, measured on two samples of Bhabha and diphoton events~\cite{Abudinen_2020}, is evaluated from the difference observed between the results from the two methods and amounts to 0.6\% relative uncertainty. 
We also assign an uncertainty of 0.003 nb on the $\tau$-pair production cross section at the c.m.~energy of 10.58~GeV, according to Ref.~\cite{Banerjee:2007is}.

\begin{table}[!htb]
  \caption{Relative systematic uncertainties entering the upper limit computation as a function of the decay mode.}
  \centering
    \begin{tabular}{cccc}
      \hline \hline
      \multirow{2}{*}{~~Affected quantity~~} & \multirow{2}{*}{~~Source~~} & \multicolumn{2}{c}{~~Mode~~}\\
       & & ~~$e\phi$~~ & ~~$\mu\phi$~~\\
      \hline
      \multirow{4}{*}{~~$\varepsilon_{\mathrm{\ell\phi}}$~~} & ~~Particle identification~~ & ~~0.8\%~~ & ~~0.3\%~~ \\
      & ~~Tracking efficiency~~ & \multicolumn{2}{c}{~~0.9\%~~} \\
      & ~~Trigger efficiency~~ & ~~0.4\%~~ & ~~0.9\%~~ \\
      & ~~Signal variable mismodeling~~ & ~~15.2\%~~ & ~~8.5\%~~ \\
      \hline 
      ~~$N_{\mathrm{exp}}$~~ & ~~Momentum scale~~ & ~~0.6\%~~ & ~~0.4\%~~ \\
      \hline
      ~~$L$~~ & ~~Luminosity~~ & \multicolumn{2}{c}{~~0.6\%~~} \\
      \hline
      ~~$\sigma_{\tau\tau}$~~ & ~~Tau-pair cross section~~ & \multicolumn{2}{c}{~~0.3\%~~} \\
      \hline \hline
    \end{tabular}
  \label{tab:syst}
\end{table}

\section{Results and conclusions}
\begin{figure}[h!]
 \begin{center}
     \includegraphics[width=0.9\textwidth]{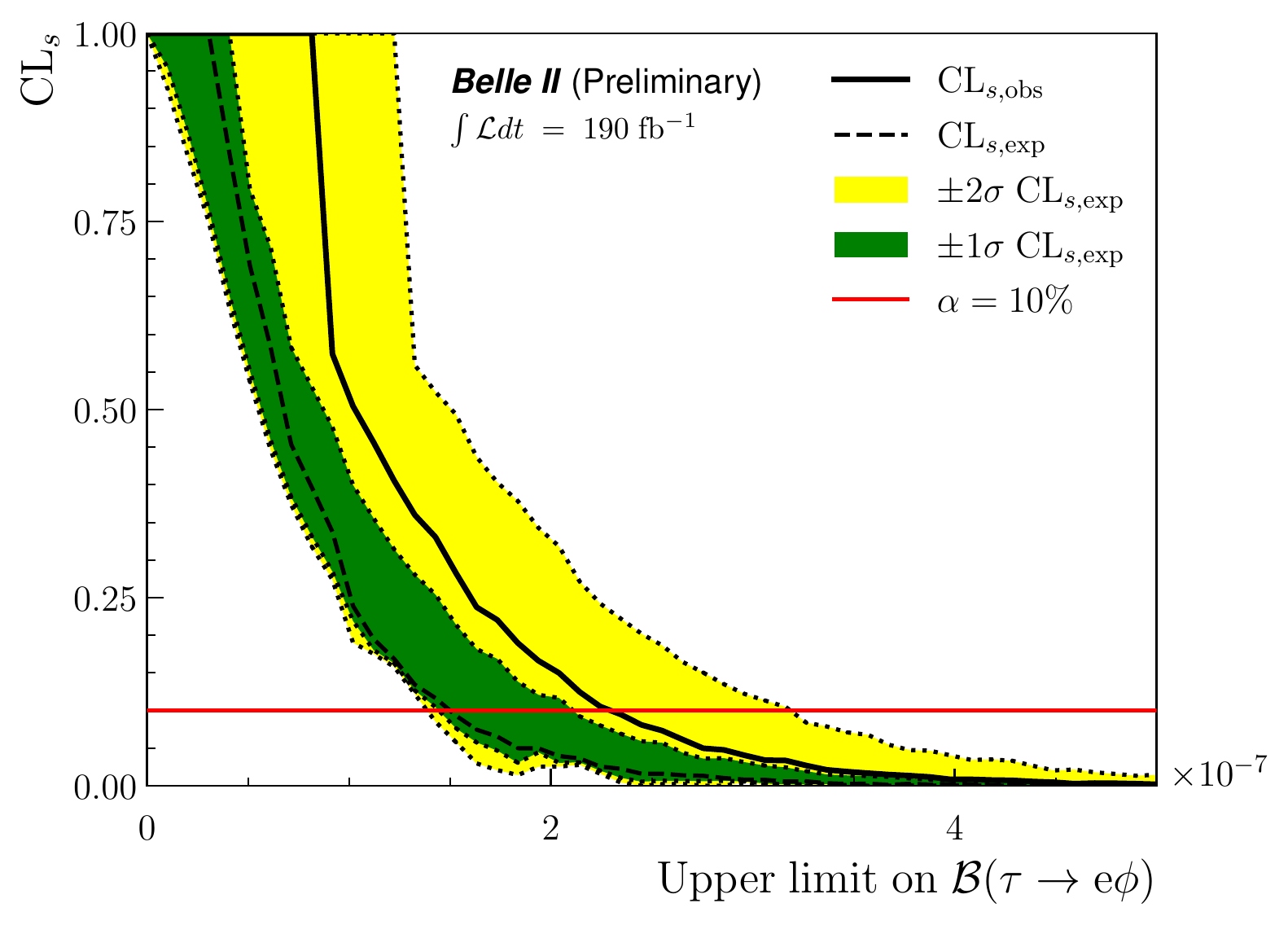}
  \hfill
      \includegraphics[width=0.9\textwidth]{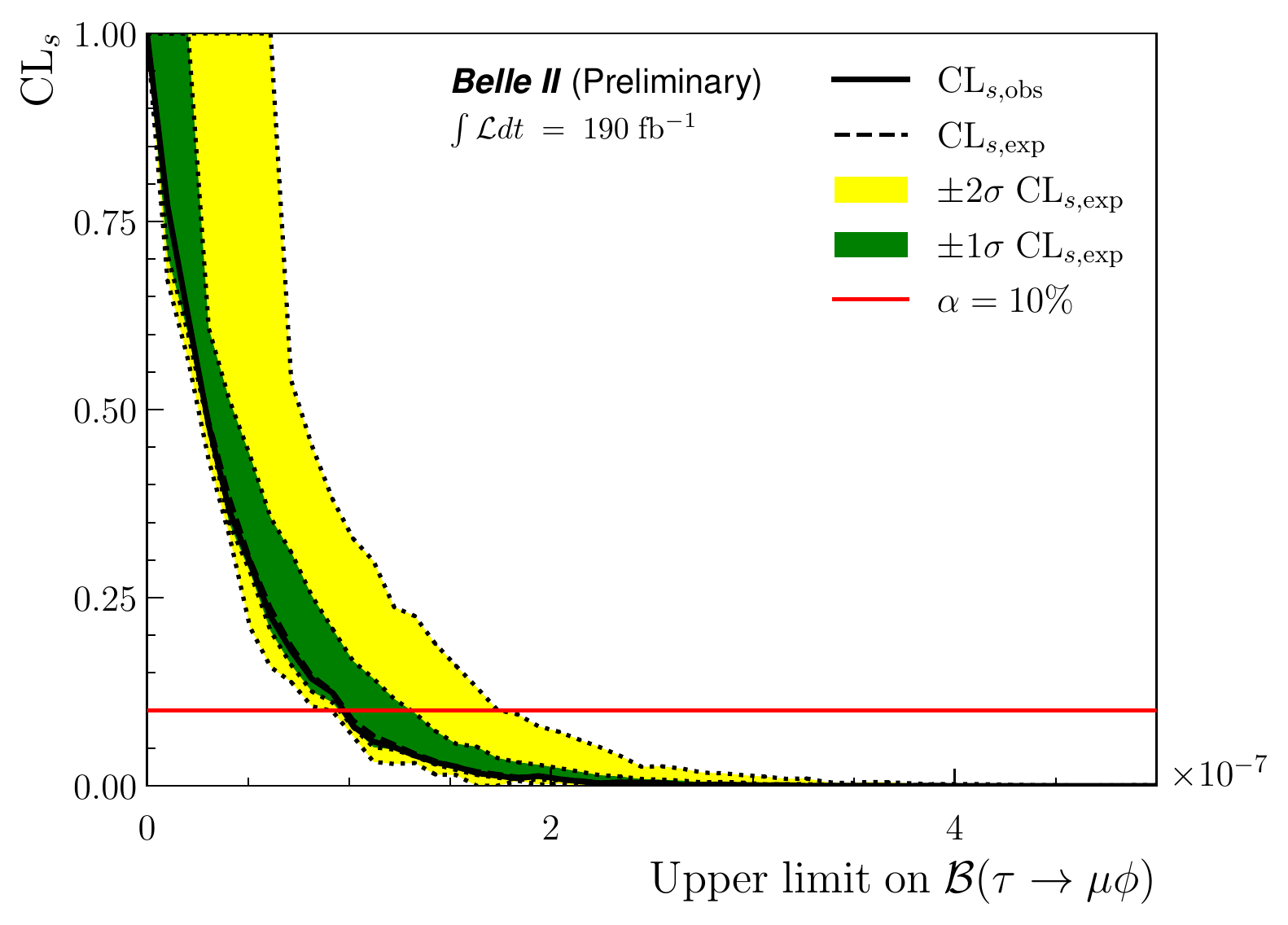}
\caption{Observed (solid black curve) and expected (dashed black curve) $\mathrm{CL}_s$ as a function of the assumed branching fractions of \ephi{} (top) and \mphi{} (bottom). The red lines correspond to the 90\% C.L.s}
\label{fig:CLs}
 \end{center} 
\end{figure}

In 190\invfem\ data, after the final selections, we observe two and zero events for the electron and muon channels, respectively, inside the $\pm 3\delta$ and $\pm 2\delta$ signal regions, as shown in the scatter plots in Figure~\ref{fig:ScPlotsData}. Data yields are compatible with expectations in Table~\ref{tab:finalRes}. Figure~\ref{fig:CLs} shows the observed $\mathrm{CL}_{s}$ curves in agreement with the expected within the yellow $\pm 2\sigma$ band. In the absence of any significant signal excess, we determine the 90\% C.L. upper limits on the branching fractions to be $\mathcal{B}(\tau^- \rightarrow e^-\phi)$ < $23\times 10^{-8}$ and $\mathcal{B}(\tau^- \rightarrow \mu^-\phi)$ < $9.7\times 10^{-8}$. 
In contrast to the one-prong tag technique used by previous searches at Babar and Belle, this analysis demonstrates the successful first application of the new untagged approach, combined with a background suppression via BDT classifiers, in the reconstruction of tau pairs events at Belle II. With this method, we obtain twice the final signal efficiency for the muon mode.
\clearpage
\FloatBarrier

\section*{Acknowledgements}
This work, based on data collected using the Belle II detector, which was built and commissioned prior to March 2019, was supported by
Science Committee of the Republic of Armenia Grant No.~20TTCG-1C010;
Australian Research Council and research Grants
No.~DE220100462,
No.~DP180102629,
No.~DP170102389,
No.~DP170102204,
No.~DP150103061,
No.~FT130100303,
No.~FT130100018,
and
No.~FT120100745;
Austrian Federal Ministry of Education, Science and Research,
Austrian Science Fund
No.~P~31361-N36
and
No.~J4625-N,
and
Horizon 2020 ERC Starting Grant No.~947006 ``InterLeptons'';
Natural Sciences and Engineering Research Council of Canada, Compute Canada and CANARIE;
Chinese Academy of Sciences and research Grant No.~QYZDJ-SSW-SLH011,
National Natural Science Foundation of China and research Grants
No.~11521505,
No.~11575017,
No.~11675166,
No.~11761141009,
No.~11705209,
and
No.~11975076,
LiaoNing Revitalization Talents Program under Contract No.~XLYC1807135,
Shanghai Pujiang Program under Grant No.~18PJ1401000,
Shandong Provincial Natural Science Foundation Project~ZR2022JQ02,
and the CAS Center for Excellence in Particle Physics (CCEPP);
the Ministry of Education, Youth, and Sports of the Czech Republic under Contract No.~LTT17020 and
Charles University Grant No.~SVV 260448 and
the Czech Science Foundation Grant No.~22-18469S;
European Research Council, Seventh Framework PIEF-GA-2013-622527,
Horizon 2020 ERC-Advanced Grants No.~267104 and No.~884719,
Horizon 2020 ERC-Consolidator Grant No.~819127,
Horizon 2020 Marie Sklodowska-Curie Grant Agreement No.~700525 "NIOBE"
and
No.~101026516,
and
Horizon 2020 Marie Sklodowska-Curie RISE project JENNIFER2 Grant Agreement No.~822070 (European grants);
L'Institut National de Physique Nucl\'{e}aire et de Physique des Particules (IN2P3) du CNRS (France);
BMBF, DFG, HGF, MPG, and AvH Foundation (Germany);
Department of Atomic Energy under Project Identification No.~RTI 4002 and Department of Science and Technology (India);
Israel Science Foundation Grant No.~2476/17,
U.S.-Israel Binational Science Foundation Grant No.~2016113, and
Israel Ministry of Science Grant No.~3-16543;
Istituto Nazionale di Fisica Nucleare and the research grants BELLE2;
Japan Society for the Promotion of Science, Grant-in-Aid for Scientific Research Grants
No.~16H03968,
No.~16H03993,
No.~16H06492,
No.~16K05323,
No.~17H01133,
No.~17H05405,
No.~18K03621,
No.~18H03710,
No.~18H05226,
No.~19H00682, 
No.~22H00144,
No.~26220706,
and
No.~26400255,
the National Institute of Informatics, and Science Information NETwork 5 (SINET5), 
and
the Ministry of Education, Culture, Sports, Science, and Technology (MEXT) of Japan;  
National Research Foundation (NRF) of Korea Grants
No.~2016R1\-D1A1B\-02012900,
No.~2018R1\-A2B\-3003643,
No.~2018R1\-A6A1A\-06024970,
No.~2018R1\-D1A1B\-07047294,
No.~2019R1\-I1A3A\-01058933,
No.~2022R1\-A2C\-1003993,
and
No.~RS-2022-00197659,
Radiation Science Research Institute,
Foreign Large-size Research Facility Application Supporting project,
the Global Science Experimental Data Hub Center of the Korea Institute of Science and Technology Information
and
KREONET/GLORIAD;
Universiti Malaya RU grant, Akademi Sains Malaysia, and Ministry of Education Malaysia;
Frontiers of Science Program Contracts
No.~FOINS-296,
No.~CB-221329,
No.~CB-236394,
No.~CB-254409,
and
No.~CB-180023, and No.~SEP-CINVESTAV research Grant No.~237 (Mexico);
the Polish Ministry of Science and Higher Education and the National Science Center;
the Ministry of Science and Higher Education of the Russian Federation,
Agreement No.~14.W03.31.0026, and
the HSE University Basic Research Program, Moscow;
University of Tabuk research Grants
No.~S-0256-1438 and No.~S-0280-1439 (Saudi Arabia);
Slovenian Research Agency and research Grants
No.~J1-9124
and
No.~P1-0135;
Agencia Estatal de Investigacion, Spain
Grant No.~RYC2020-029875-I
and
Generalitat Valenciana, Spain
Grant No.~CIDEGENT/2018/020
Ministry of Science and Technology and research Grants
No.~MOST106-2112-M-002-005-MY3
and
No.~MOST107-2119-M-002-035-MY3,
and the Ministry of Education (Taiwan);
Thailand Center of Excellence in Physics;
TUBITAK ULAKBIM (Turkey);
National Research Foundation of Ukraine, project No.~2020.02/0257,
and
Ministry of Education and Science of Ukraine;
the U.S. National Science Foundation and research Grants
No.~PHY-1913789 
and
No.~PHY-2111604, 
and the U.S. Department of Energy and research Awards
No.~DE-AC06-76RLO1830, 
No.~DE-SC0007983, 
No.~DE-SC0009824, 
No.~DE-SC0009973, 
No.~DE-SC0010007, 
No.~DE-SC0010073, 
No.~DE-SC0010118, 
No.~DE-SC0010504, 
No.~DE-SC0011784, 
No.~DE-SC0012704, 
No.~DE-SC0019230, 
No.~DE-SC0021274, 
No.~DE-SC0022350; 
and
the Vietnam Academy of Science and Technology (VAST) under Grant No.~DL0000.05/21-23.

These acknowledgements are not to be interpreted as an endorsement of any statement made
by any of our institutes, funding agencies, governments, or their representatives.

We thank the SuperKEKB team for delivering high-luminosity collisions;
the KEK cryogenics group for the efficient operation of the detector solenoid magnet;
the KEK computer group and the NII for on-site computing support and SINET6 network support;
and the raw-data centers at BNL, DESY, GridKa, IN2P3, INFN, and the University of Victoria for offsite computing support.

\printbibliography[heading=bibintoc]

\clearpage

\end{document}